\def    \S      {\Sigma}

\def    \to     {\rightarrow}

\def    \D      {\Delta}

\def    \a      {\alpha}

\def    \A      {{\cal A}}

\def    \R      {{\cal R}}

\def    \se     {\subseteq}

\def    \lcc#1  {\langle\!\langle #1 \rangle\!\rangle}
\def    \rcc#1  {\langle #1 \rangle}
\def    \cc#1   {[#1]}

\def    \ft#1   {\footnote{#1} }

\def	\nodeq#1 {\stackrel{#1}{\sim}}
\def	\trans#1 {\stackrel{#1}{\longrightarrow}}

\def	\ds     {\displaystyle }
\def	\comment#1 {}

\def    \comm#1  {\begin{center} {\tt $<<<$ #1 $>>>$ } \end{center}}

\def\GF{\hbox{\raise 1.6pt \hbox{$\varphi$}}}

\documentclass[submission,copyright,creativecommons]{eptcs}
\providecommand{\event}{HSB 2013}
\usepackage{amssymb,amsmath,times,amsfonts,times,color,graphicx,latexsym}

\title{Exploring the Dynamics of Mass Action Systems}

\author{Oded Maler\thanks{Supported by the ANR projects Syne2Arti, Eqinocs and Cadmidia.} \institute{CNRS-VERIMAG\\ University of Grenoble} \email{Oded.Maler@imag.fr}
\and \'Ad\'am M.\ Hal\'asz\thanks{Supported by NIHGrant K25 CA131558. Work done while the author was visiting  CNRS-VERIMAG.} \institute{Department of Mathematics\\ West Virginia University} \email{halasz@math.wvu.edu}
\and Olivier Lebeltel  \institute{CNRS-VERIMAG\\ University of Grenoble} \email{Olivier.Lebeltel@imag.fr}
\and Ouri Maler \institute{Grenoble} \email{ouri.maler@gmail.com}}
\def\titlerunning{Mass Action Dynamics}
\def\authorrunning{O. Maler, A. H\'alasz, O. Lebeltel, O. Maler}

\begin{document}
\newtheorem{definition}{Definition}
\newtheorem{theorem}{Theorem}
\newtheorem{corollary}{Corollary}
\newtheorem{observation}{Observation}
\newtheorem{algorithm}{Algorithm}

\newcommand{\N}{\mathbb{N}}
\renewcommand{\R}{\mathbb{R}}
\newcommand{\Z}{\mathbb{Z}}
\newcommand{\B}{\mathbb{B}}

\newcommand{\op}{\sqcap}
\newcommand{\popu}{{\bf Populus}}

\maketitle

\comment{
\begin{abstract}
We develop the \emph{Populus} toolkit for exploring the dynamics of mass action systems under different assumptions. To simplify the model and focus on what we believe to be the essence of the mass action phenomena, we make some simplifying assumptions that can be eventually relaxed.
\end{abstract}
}

\section{Introduction}

\emph{Mass action} is a fundamental notion in many situations in Chemistry, Biochemistry, Population Dynamics and Social Systems \cite{ball2004critical}. In this class of phenomena, one has a large population of individuals partitioned into several types of ``species'', whose dynamics is specified by a set of reaction rules. Each reaction indicates the transformation that is likely to take place when individuals of specific types come into contact. For example, a rule of the form $~~A+B~~\to~~ A+C~~$ says that each time an instance of $A$ meets an instance of $B$, the latter is transformed into a $C$. Denoting by $n_A$ and $n_B$ the number of instances of $A$ and $B$ existing at a certain moment, the likelihood of an $(A,B)$-encounter is proportional to $n_A\cdot n_B$. Hence the rate of change of $n_B$ will have a \emph{negative} contribution proportional to $n_A\cdot n_B$ and that of $n_C$ will have the same magnitude of \emph{positive} contribution. Combining for each of the species the negative contributions due to reactions in which it is transformed into something else  with the positive contributions due to reactions that yield new instances of it, one typically obtains a system of \emph{polynomial}\ft{Actually \emph{bilinear} if one assumes the probability of triple encounters to be zero, as is often done in Chemistry.} differential/difference equations.

The goal of the research program sketched in this paper (initially inspired by \cite{cardelli2009artificial}) is to build a class of \emph{synthetic} mathematical models for such systems, admitting some nice and clean properties which will reflect essential and fundamental aspects of mass action behavior while  at the same time abstract away from accidental real-life details due to Chemistry, Physics and even some Geometry. Introducing such details at this preliminary stage would obscure the essence and render the analysis more complex. These models will then be subject to various investigations by analytical, simulation-based and other methods to  explore their dynamics and discover the principles that govern their behavior. Such investigations may eventually lead to novel ways to control mass action systems with potential applications, among others, in drug design and social engineering. These issues have been studied, of course, for many years in various contexts and diverse disciplines, \cite{leboudec2007,bortolussi2013} to mention a few, but we hope, nevertheless, to provide a fresh look on the subject.


The rest of this paper  is organized as follows. In Section~2 we present the basic model of the individual agent (particle) as a \emph{probabilistic automaton} capable of being in one out of several states, and where transition labels refer to the state of the agent it encounters at a given moment. We then discuss several ways to embed these individual agents in a model depicting the evolution of a large ensemble of their instances. In Section~3 we describe three such aggregate models. We start with a rather standard model where state variables correspond to the relative concentrations of agent types. Such models depict the dynamics of the \emph{average} over all behaviors and they  are traditionally ODEs but we prefer to work in discrete time to simplify the notation. The second model is based on stochastic simulation under the well-stirred assumption with no modeling of \emph{space}, which is introduced in the third model where particles wander in space in some kind of random motion and a reaction takes place when the distance between two particles becomes sufficiently small. The model thus obtained is essentially a kind of a \emph{reaction-diffusion} model. In Section~4  we briefly describe the {\bf Populus} tool kit that we developed for exploring the dynamics of such models and illustrate its functionality by demonstrating some effects of the initial spatial distributions of some particles that lead to deviation from the predictions of a well-stirred model.

\section{Individual Models and Aggregation Styles}

We consider mass action systems where new individuals are not born and existing ones neither die nor aggregate into compound entities: they only change their \emph{state}.

\subsection{Individuals}

A particle can be in one of finitely-many states and its (probabilistic) dynamics depicts what happens to it (every time instant) either spontaneously or upon encountering another particle. The  object specifying a particle is a probabilistic automaton:\ft{A probabilistic automaton \cite{paz} is a Markov chain with an \comment{external} input alphabet where each input symbol induces a different transition matrix. It is called sometimes a Markov Decision Process  (MDP) but we prefer to reserve this term for a strategy synthesis problem in a game where the alphabet denotes the controller's action against a stochastic adversary. In the present model there is no ``decision" associated with the input as it is an external (to the particle) influence on the dynamics.}

\begin{definition}[Probabilistic Automaton]
A probabilistic automaton is a triple $\A=(Q,\S,\delta)$ where $Q$ is a finite set of states, $\S$ is a finite  input alphabet and $\delta:Q\times \S \times Q\to \R$ is a probabilistic transition function such that for every $q\in Q$ and $a\in \S$,
$$\sum_{q'\in Q}\delta(q,a,q')=1.$$
\end{definition}
In our model $Q=\{q_1,\ldots, q_n\}$ is the set of particle types and each instance of the automaton is always in one of those. The input alphabet is $Q\cup\{\bot\}$ intended to denote the type of \emph{another} particle encountered by the automaton and with the special symbol $\bot$ indicating a non-encounter. Intuitively, $\delta(q_1,q_2,q_3)$ represents the probability that an agent of type $q_1$ converts to type $q_3$ when it encounters an agent of type $q_2$. Likewise $\delta(q_1,\bot,q_3)$ is the probability of becoming $q_3$ spontaneously without meeting anybody. Table~\ref{tab:prob-aut} depicts a $3$-species probabilistic automaton. We use the notation $q_1 \trans{q_2} q_3$ for an actual invocation of the rule, that is, drawing an element of $Q$ according to probability $\delta(q_1,q_2,.)$ and obtaining $q_3$ as an outcome.

 In general our models are \emph{synchronous} with respect to time: time evolves in fixed-size  steps and at every step each particle detects whether it encounters another (and of what type) and takes the appropriate transition.  The interpretation of when an agent meets another depends, as we shall see, on additional assumptions on the global aggregate model. It is worth noting that we restrict ourselves here to reaction rules which are ``causal" in the following sense: when an $(A,B)$-encounter takes place, the influence of $A$ on $B$  and the influence of $B$ on $A$ are \emph{independent}. Hence not all types of probabilistic rewrite rules of the form
$~~A+B \to A_1+B_1~(p_1)~|~ A_2+B_2~(p_2)~|~\cdots~|~ A_k+B_k~(p_k)~~ $
can be realized, only those that are products of simple rules. This restriction is not crucial for our approach but it simplifies some calculations.

\comment{
Sometimes, it might be useful to consider further restrictions, for example that an agent does not change when it meets someone of the same type: $\delta (q,q,q)=1$ and $\delta (q,q,q')=0$ when $q'\not = q$.  This prevents squaring in the dynamic law and keeps the system multi-linear, a fact that can facilitate root isolation.
}

\begin{table}[h]
$$
\begin{array}{c}
\begin{array}{|l||lll||lll|lll|lll|}
\hline
\delta    &    \multicolumn{3}{c}{\bot} & \multicolumn{3}{c}{q_1}  & \multicolumn{3}{c}{q_2} & \multicolumn{3}{c}{q_3} \vline \\
\hline
q_1 &  0.9 &  0.1 & 0.0 & 1.0 & 0.0 & 0.0 & 0.7 & 0.2 & 0.1 & 0.7 & 0.0 & 0.3 \\
q_2 &  0.1 &  0.8 & 0.1 & 0.0 & 0.6 & 0.4 & 0.0 & 1.0 & 0.0 & 0.1 & 0.9 & 0.0 \\
q_3 &  0.0 &  0.0 & 1.0 & 0.7 & 0.0 & 0.3 & 0.3 & 0.4 & 0.3 & 0.0 & 0.0 & 1.0 \\
\hline
\end{array}
\\ ~\\~\\
\begin{array}{|l|}
\hline
x'_1 = x_1 -0.09 x_1 +0.09 x_2 -0.06x_1 x_2 +0.08 x_1 x_3 +0.08 x_2x_3\\
x'_2 = x_2 + 0.09 x_1 -0.18 x_2-0.04x_1 x_2+0.06 x_2 x_3\\
x'_3 = x_3 + 0.09 x_2 + 0.1x_1 x_2 -0.08 x_1 x_3 -0.14 x_2 x_3\\
\hline
\end{array}
\\ ~\\~\\
\begin{array}{|l|}
\hline
x'_1 = x_1 -0.01 x_1+0.01 x_2-0.54x_1 x_2+0.72 x_1 x_3 +0.72 x_2x_3\\
x'_2 = x_2 + 0.01 x_1 -0.02 x_2-0.36 x_1 x_2+0.54 x_2 x_3\\
x'_3 = x_3 +0.01 x_2  +0.9x_1 x_2 -0.72 x_1 x_3 -1.26 x_2 x_3\\
\hline
\end{array}
\end{array}
$$

\caption{\label{tab:prob-aut} A $3$-species probabilistic automaton, and the average dynamical system derived for the sparse situation $\a=0.1$ and for the dense situation $\a=0.9$. Starting from initial state $x=(0.4,0.3,0.3)$ the first system converges to the state
$(0.366,0.195,0.437)$ while the second converges to $(0.939 ,0.027, 0.033)$.}
\end{table}

\subsection{Aggregation Styles}

Consider now a set $S$ consisting of $m$ individuals  put together, each being modeled as an automaton. A global configuration of such a system should specify, at least, the state of each particle, resulting in the enormous state space $Q^S$ consisting of $n^m$ states (micro-states in Physpeak). A very useful and commonly-used abstraction is the \emph{counting abstraction} obtained by considering two micro-state equivalent if they agree on the number of  particles of each type, regardless of their particular identity. The equivalence classes of this relation form an abstract state-space of macro-states (also known as particle count representation) $P\se S^Q$ consisting  of $n$-dimensional vectors:
  $$P= \{(X_1,\ldots,X_n):\forall i~0\leq X_i \leq m \wedge \sum_{i=1}^n X_i= m\}.$$
 The formulation of a model that tracks the evolution of an ensemble of particles can be done in different styles. For our purposes we classify models according to two features: 1) Individual vs.\ average dynamics and 2) Spatially-extended vs.\ well-stirred dynamics. These two features are related but not identical.

 For the first point, let us recall the trivial but important fact that we have a non-deterministic system where being in a given micro-state, each particle tosses  one or more coins, properly biased according to the states of the other particles, so as to determine its next state. To illustrate, consider a rule which transforms a particle type $A$  into $B$ with probability $p$. Starting with $m$ instances of $A$, there will be $m$ coin tosses each with probability $p$ leading to some number close to $m\cdot p$ indicating how many $A$'s  convert into $B's$. Each individual run will yield a different number (and a different sequence of subsequent numbers) but on the average (over all runs) the number of $A$'s will be reduced in the first step from $m$ to $m\cdot(1- p)$.
 
  \emph{Individualistic} models, that is, stochastic simulation algorithms (SSA),  generate such runs, one at the time. On the other hand, ``deterministic" ODE models compute et every step the average number of particles for each type where this average is taken (in parallel) over all runs . For well-behaving systems, the relationship between this averaged trajectory and individual runs is of great similarity: the evolution in actual runs will appear as fluctuating around the evolution of the average. \comment{ (see Fig.~\ref{fig:average-indiv}-(a)).}  On the other hand, when we deal with more complex systems where, for example, trajectories can switch into two or more distinct and well-separated equilibria, the behavior of the average is not so informative. \comment{ as shown in Fig.~\ref{fig:average-indiv}-(b).} There is a whole research thread, starting with \cite{gillespie}, that feeds on this important distinction (see \cite{deviant,julius2008stochastic} for further discussions).

\comment{
\begin{figure}
  \centering
\begin{tabular}{ccc}
  \includegraphics[width=7cm]{average-indiv-simple.jpg} & &
    \includegraphics[width=7cm]{average-indiv-complex.jpg} \\
    (a) & & (b)
  \end{tabular}
  \caption{(a) The evolution of a simple system where $A$ is eventually transformed to $B$. The smoother curves depict the average while the other show an individual trajectory. (b) A system where the average stabilizes while individual trajectories fluctuate. }\label{fig:average-indiv}
\end{figure}
}

 The other issue is whether and how one models the distribution of particles in \emph{space}. Ignoring the spatial coordinates of particles, the probability of a particular transition being taken depends only on the total \emph{numbers} of particles of each type, which is equivalent to the \emph{well-stirred} assumption: all instances of each particle type are distributed uniformly in space and hence all particles will see the same proportion of other particles in their neighborhood.  On the other hand, in spatially extended models each particle is endowed with a location which changes quasi-randomly and what it encounters in its moving neighborhood determines the interactions it is likely to participate in. \comment{Hence a moving particle will be exposed more to what happens locally along its spatial trajectory than to the average number of particles.}


\section{Implemented Aggregate Models}

In the sequel we describe in some detail the derivation of three models: average dynamics, individual well-stirred dynamics and spatially-extended dynamics. All our models are in discrete time which will hopefully make them more accessible to those for whom the language of integrals is not native. For the others, note that our model corresponds to a fixed time-step simulation of ODEs.

\subsection{Average Well-Stirred Dynamics}

To develop the average dynamics under the well-stirred assumption.\ft{Using PDEs one can sometimes derive average models under \emph{non-uniform} distributions of particles in space but most chemical reaction models employ the well-stirred assumption.} we normalize the global macro-state of the system, a vector $X=(X_1,\ldots,X_n)$, into  $x=(x_1,\ldots,x_n)$ with $x_i=X_i/m$ and hence $\sum x_i=1$ (relative concentration). Let  $\a$, $0\leq \a\leq 1$ be a density parameter which determines the probability of bumping into another particle in one time step. The evolution in this state space over time is the outcome of playing the following protocol at every time step. First,  $\a S$ of the particles bump into others and hence follow a binary reaction rule while the remaining $(1-\a)S$ particles do not interact and hence follow the solitary transition function. We will derive the dynamics, which is of the general form\ft{We use the primed notation where $x$ stands for $x[t]$ and $x'$ denotes $x[t+1]$.}
~$x'=x+\D(x)$. For each variable, the additive change can be written as
$$
\begin{array}{c}
\ds{ \D(x_k)=(1-\a)\D_1(x_k)+ \a \D_2(x_k) }\\
\ds{\D_1(x_k)=\sum_{i=1}^n( x_i \cdot \delta(x_i,\bot,x_k) - x_k\cdot \delta(x_k,\bot,x_i)) }\\
\ds{\D_2(x_k)=\sum_{i=1}^n \sum_{j=1}^n  ( x_i x_j \cdot \delta(x_i,x_j,x_k) - x_k x_i \cdot \delta(x_k,x_i,x_j))}
\end{array}
 $$
 Here,  $\D_1$ and $\D_2$ are the expected contributions to $x_k$ by the solitary (resp.\ binary) reactions, each
summing up the transformations of other agents into type $k$ minus the transformation of type $k$ into other types.
Thus, we obtain a discrete-time bilinear dynamical system, which is linear when  $\a=0$, see example in  Table~\ref{tab:prob-aut}. As already mentioned, this deterministic dynamics tracks the evolution of the \emph{average} concentration of particles over all individual runs.

\comment{
$$D_2(x_k)=\sum_{i=1}^n \sum_{j=1}^n x_i x_j C_2((x_i,x_j),x_k)/2$$
where $C_1(x_i,x_k)$ is the expected contribution of one instance of an agent of type $i$ following a solitary rule:
$$
C_1(x_i,x_k)=p (q_i,\bot,q_k) -\delta_{ik}
$$
and $C_2((x_i,x_j),x_k)$ is the expected contribution to $x_k$ of a single encounter between species of type $(i,j)$:
$$
C_2((x_i,x_j),x_k)= p (q_i,q_j,q_k) + p (q_j,q_i,q_k)  -\delta_{ik} -\delta_{jk}
$$
where $\delta_{ij}=1$ when $i=j$ and $\delta_{ij}=0$ otherwise.\ft{Not that the division by $2$ in the definition of $D_2$ is due to the fact that $N$ particales can participate in $N/2$ binary reactions.}
In other words, $C_1(x_i,x_k)$ is the expected change in the number of $q_k$ instances if we pick one (solitary) $q_i$ and let it behave according to the transition table. If $i=j$ this is a non-positive contribution, of course. Likewise $C_2((x_i,x_j),x_k)$ is the expected change in the number of instances of $q_k$ if we pick a pair consisting of one $q_i$ and one $q_j$ and let them collide.
}

\subsection{Individual Well-Stirred Dynamics}

The second model, whose average behavior is captured by the previous one, generates individual behaviors without spatial information. A micro-state of the systems is represented as a set $L$ of particles, each denoted as $(g,q)$ where $g$ is the particle identifier and $q$ is its current state.

\begin{algorithm}[Individual Well-Stirred Dynamics]
~\\
\begin{tabular}{l}
{\bf Input}: A list $L$ of particles and states \\
{\bf Output} A list $L'$ representing the next micro-state \\
~~\\
$L':=\emptyset$\\
{\bf repeat}\\
~~~{\bf draw} a random particle $(g,q) \in L$; $L:=L-\{(g,q)\}$\\
~~~{\bf draw} binary/solitary with probability $\a$\\
~~~{\bf if} solitary {\bf then} \\
~~~~~~apply solitary rule $q\trans{\bot} q'$\\
~~~~~~$L':=L'\cup\{(g,q')\}$\\
~~~{\bf else} \\
~~~~~~{\bf draw} a random particle $(g',q')\in L$; $L:=L-\{(g',q')\}$\\
~~~~~~apply binary rules $q\trans{q'} q''$ and $q' \trans{q} q'''$\\
~~~~~~$L':=L'\cup\{(g,q''),(g',q''')\}$\\
~~~{\bf endif}\\
{\bf until} $L=\emptyset$
\end{tabular}
\end{algorithm}
After each update round, particle types are counted to create macro-states. The algorithm can most likely be made more efficient  by working directly on macro-states and drawing the increments of each particle type using a kind of binomial distribution that sums up the multiple coin tosses. Similar ideas underlie the $\tau$-leaping algorithm of \cite{gillespie2001approximate}.

\comment{We first explain this model procedurally. Suppose we have $S$ particles and we have one state variable for each, indicating its type.
At each time step, each of those will draw one or more coins and decide how to change its state. The first coins will decide whether it meets another particle according to the density. If a collision is chosen, the type of the other will be determined according to the current global number of each type. Repeating this for every round for all particles, we have individual trajectories of the total number of particles of each type. I believe that this model can be transformed into a stochastic dynamical system over the concentration (without individuals) roughly of the form $x_{t+1}+\D(x) + u$ where $\D(x)$ is the average contribution and $u$ is some noise, distributed normally around zero, reflecting the independent coin tosses made by instances of $x$.
}

\comment{
 Unlike the averaged model mentioned above having one state variable for each type, a system of $S$ particles will have at least $S$ state variables indicating the state of each and every particle.  The difference between non-spatial and spatial is the following: in the former, the choice of what a particle of type $A$ will do in the next step , so with some probability it will choose whether or not to meet someone else and, if it meets someone else, the identity of the latter is determined by throwing a coin according to the proportions of $n_A$ , $n_B$, etc. Since everybody is assumed independent, maybe there is a simpler procedure which works on total numbers and computes the next value of $n_A$ as distributed normally around $n_A+\D$ where $\D$ is the average contribution. The other type of individual model is
}

\subsection{Individual Spatial Dynamics}

Our third model does take space into account by representing each particle as $(g,q,y)$ with $y$ being it spatial coordinates, currently ranging over a bounded rectangle. The next state is computed in two phases that correspond to diffusion and reaction. First, each particle is displaced by a vector of random direction and magnitude (bounded by a constant $s$). For mathematical convenience reasons we use \emph{periodic boundary conditions} so that when a particle crosses the boundary of the rectangle it reappears on the other side as if it was a torus. Then for each particle we compute its set of neighbors $N$, those  residing in a ball of a  pre-specified {\em interaction radius} $r$, typically in the same order of magnitude as $s$. If the particle has several neighbors\ft{Which turns out not to be negligible with the parameters we have chosen so far which are unlike commonly-used models where the average distance between particles is orders of magnitude larger than the interaction radius $r$.} we compute the outcome of all those possible interactions and choose among them randomly.

\begin{algorithm}[Individual Spatial Dynamics]
~\\
\begin{tabular}{l}
{\bf Input}: A list $L$ of particles and states including planar coordinates \\
{\bf Output} A list $L'$ representing the next micro-state \\
~~\\
$L':=\emptyset$\\
{\bf foreach} particle $(g,q,y)\in L$\\
~~~{\bf draw} randomly $h\in[0,s]$ and  $\theta\in[0,2\pi] $ \\
~~~$y:=y+(h,\theta)$ \\
{\bf endfor} \\
{\bf foreach} $(g,q,y)\in L$\\
~~~$N:=\{(g',q',y'):d(y,y')<r\}$\\
~~~{\bf if} $N=\emptyset$ {\bf then}\\
~~~~~~apply solitary rule $q\trans{\bot} q'$\\
~~~~~~$L':=L'\cup\{(g,q',y)\}$\\
~~~{\bf else} \\
~~~~~~$M:=\emptyset$\\
~~~~~~{\bf foreach} $(g',q',y')\in N$\\
~~~~~~~~~apply binary rule $q\trans{q'} q''$\\
~~~~~~~~~$M:=M\cup \{q''\}$\\
~~~~~~{\bf endfor}\\
~~~~~~{\bf draw} $q'\in M$\\
~~~~~~$L':=L'\cup\{(g,q',y)\}$\\
~~~{\bf endif}\\
{\bf endfor}
\end{tabular}
\end{algorithm}
The connection between this model, embedded in a rectangle of area $W$, and the non-spatial ones is made via the computation of the density factor $\a$. The probability of a particle \emph{not} interacting with another particle is its probability to be outside its interaction ball, that is, $1-\frac{\pi r ^2}{W}$,  and the odds of not interacting with any of the other $m-1$ particles is $(1-\frac{\pi r ^2}{W})^{m-1}$ and hence $\a=1- (1-\frac{\pi r ^2}{W})^{m-1}$.

\section{The Populus Toolkit: Preliminary Experiments}

We developed a  prototype tool called \popu, written in Java and Swing, for exploring such dynamics. The input to the tool  is a particle automaton along with additional parameters such as the dimensions of the rectangle where particles live, the  geometric step size $s$, the interaction radius $r$ and the initial number of each particle type, possibly restricted to some sub rectangles. The tool simulates the three models, plotting the evolution of particle counts over time as well as animating the spatial evolution.

\begin{figure}
  \centering
\begin{tabular}{ccccc}
  \includegraphics[width=4.6cm]{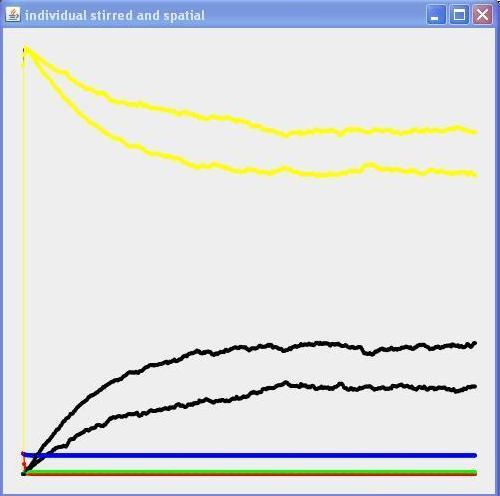} & &
  \includegraphics[width=4.6cm]{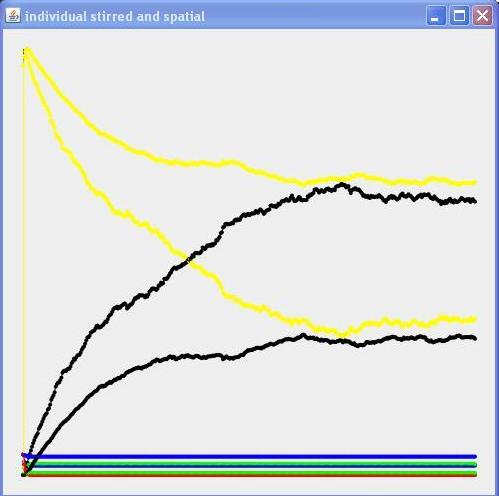} & &
  \includegraphics[width=4.6cm]{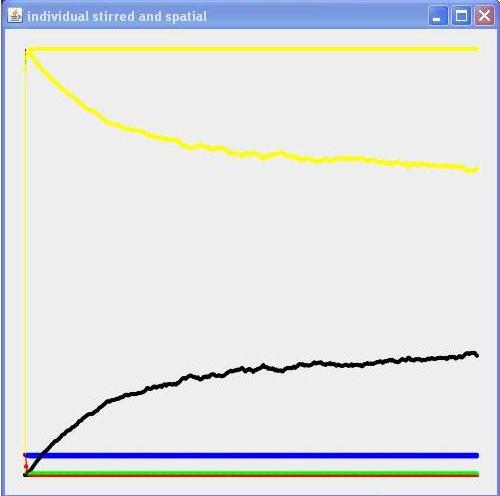}  \\
    (a) & & (b) && (c)
  \end{tabular}
  \caption{ The evolution of the $5$-species system where $A$'s and $B$'s are initially  (a) distributed uniformly in space; (b) close to each other and (c) remote from each other. The plot depict the spatial and non-spatial models with the black curve indicating the growth of $E$.}\label{fig:exp}
\end{figure}

To demonstrate the difference between spatial and non-spatial models we simulated a system with $5$ species,  $A$, $B$, $C$, $D$ and $E$ to $E$. $A$ and $B$ are initially present in small quantities, $50$ each, while $D$ has $1000$ instances. When $A$ and $B$ meet, $A$ is transformed into an active and stable agent $C$ which converts $D$'s  to $E$'s . Since $E$ is also rather stable, the emergence of $C$ will eventually convert a large number of $D$'s to $E$'s. However, $B$ is very unstable and each step it may change with probability $0.5$. Hence the initial spatial distribution of $A$ and $B$ may influence the evolution significantly. We simulated the corresponding spatial model on a $20\times 20$ square, starting from three different initial micro-states in all of which $D$ is distributed uniformly over all the square: (a) $A$ and $B$ are distributed uniformly all over space; (b) Both $A$ and $B$ are concentrated in a unit square in the middle; (c) $A$ and $B$ are concentrated inside distinct unit squares far apart from each other. The results are plotted in Figure~\ref{fig:exp}. As a first observation, in scenario~(a) the spatial model converts $D$ to $E$ slower than the well-stirred one, despite the well-stirred initial condition. It is too early to speculate about the reasons but it might be that with our parameters where reaction is not slower than diffusion, a $C$ particle converts the $D$'s in its neighborhood and hence meets less of them than what their global concentration would imply. In scenario~(b), due to the proximity of $A$ and $B$ there is a burst of $C$'s at the beginning and the spatial model progresses faster than the well-stirred. Finally, when $A$ and $B$ are initially far apart, no $C$ and hence no $E$ are produced, unlike the prediction of the well-stirred model. In all those experiments the behavior of the well-stirred model was close to that of the average model.

\vspace{-0.2cm}
\bibliographystyle{eptcs}
\bibliography{bib-populus}

\end{document}